# Nanostructured-membrane electron phase plates

Yujia Yang,[1] Chung-Soo Kim,[1] Richard G. Hobbs,[1] Phillip D. Keathley,[1] Karl K. Berggren[1]

[1]*Research Laboratory of Electronics, Massachusetts Institute of Technology, Cambridge, Massachusetts 02139, United States*

**Abstract**

Electron beams can acquire designed phase modulations by passing through nanostructured material phase plates. These phase modulations enable electron wavefront shaping and benefit electron microscopy, spectroscopy, lithography, and interferometry. However, in the fabrication of electron phase plates, the typically used focused-ion-beam-milling method limits the fabrication throughput and hence the active area of the phase plates. Here, we fabricated large-area electron phase plates with electron-beam lithography and reactive-ion-etching. The phase plates are characterized by electron diffraction in transmission electron microscopes with various electron energies, as well as diffractive imaging in a scanning electron microscope. We found the phase plates could produce a null in the center of the bright-field based on coherent interference of diffractive beams. Our work adds capabilities to the fabrication of electron phase plates. The nullification of the direct beam and the tunable diffraction efficiency demonstrated here also paves the way towards novel dark-field electron-microscopy techniques and tunable electron phase plates.

**Main**

Electron beams are commonly used in microscopy, spectroscopy, and lithography instruments. Electron wavefront engineering and beam shaping are of vital importance in these applications. Recently, phase plates made from nanostructured electron-transparent membranes have been demonstrated to impose a rich set of phase profiles in electron beams, including electron vortex beams[1–4], Airy beams[5], helicon beams[6], and arbitrarily sculptured wavefronts[7]. In this work, we fabricate large-area mesh-shaped electron phase plates with electron-beam lithography, and characterize the phase plates with electron diffraction in transmission electron microscopes (TEMs) and diffractive imaging in a scanning electron microscope (SEM). Specifically, we find the phase plates create a null in the center of the bright-field for certain electron energy.

Electron-beam lithography[6] (EBL) is an attractive alternative to the more commonly used focused ion beam (FIB) milling[1,3–5,7,8], in the fabrication of electron phase plates. Lithographic pattern definition with EBL, followed by a parallel pattern transfer step, enables high-throughput production of small-feature-size, large-area phase plates favored in shaping electron-beam probes with a tight focus. The fabricated phase plates are also free from ion-implantation-induced chemical and physical changes in the materials, which could cause variations in the phase plates[9]. Additive manufacturing of electron phase plates with EBL and a metal lift-off process was demonstrated in earlier work[6] (the fabricated structure was treated as an amplitude mask in the work). In our work, we add capabilities to the EBL fabrication of electron phase plates by demonstrating subtractive manufacturing with a dry-etching step. Using an intermediate resist layer and a follow-up pattern transfer step could enable the patterning of various materials, including those that are hard to mill, and also permit both additive and subtractive pattering techniques to be combined, enabling the production of phase plates with diverse materials and structures. Aside from electron phase plates, these nano-patterned membrane structures could also find a wide range of applications including diffractive elements for atoms[10], molecules[11], and X-rays[12], nanophotonic devices based on photonic crystals[13], separation processes based on nanoporous membranes[14], and nanopore single molecule sensing[15].

We have also observed that, for certain electron energy and hence its wavelength, the diffractive phase plates produce a null in the center of the bright-field based on coherent interference of diffractive beams, which could be considered as an inverse process of the Arago-Poisson spot with electrons[16]. The phase plates that use coherent interference to nullify the central beam could increase the image intensity, boost the signal-to-noise ratio, and improve the contrast in dark-field electron-microscopy techniques. For instance, dark-field imaging and holography techniques have been used to analyze strain and defects in crystalline samples[17,18], and to improve the image contrast in scanning transmission electron microscopy[19]. With the help of grating phase plates that diffract the incident beam and nullify the direct beam, this analysis could be performed in the bright-field with less aberration, higher image intensity, and higher signal-to-noise ratio. Additionally, no sample tilt or central beam stop would be required. Moreover, the phase plates create two-dimensional electron diffraction patterns with a null in the center (annular-shaped electron beam). This beam-shaping capability is similar to the holographic vortex-beam generation[1,3], except that only the direct beam instead of a diffracted beam is needed. The shaped electron beam could be used as a probe that effectively performs hollow-cone (or multi-beam-tilt) dark-field illumination, which improves the spatial resolution[20,21], increases the image contrast[22,23], suppresses the phase-contrast noise[24], and potentially outruns radiation damage in imaging biological samples[25].

In this work, we consider a simple mesh phase plate as schematically illustrated in Figure 1. The mesh consists of a two-dimensional array of nano-holes patterned in a thin membrane. This mesh phase plate is essentially a two-dimensional grating for electron beams. An electron beam passing through the phase plate will be diffracted into multiple diffraction orders, with the direct beam (blue) and first-order diffracted beams (red) highlighted in Figure 1. The greyscale image at the bottom of Figure 1 is an experimental electron diffraction pattern from a nanofabricated mesh phase plate measured in a transmission electron microscope.

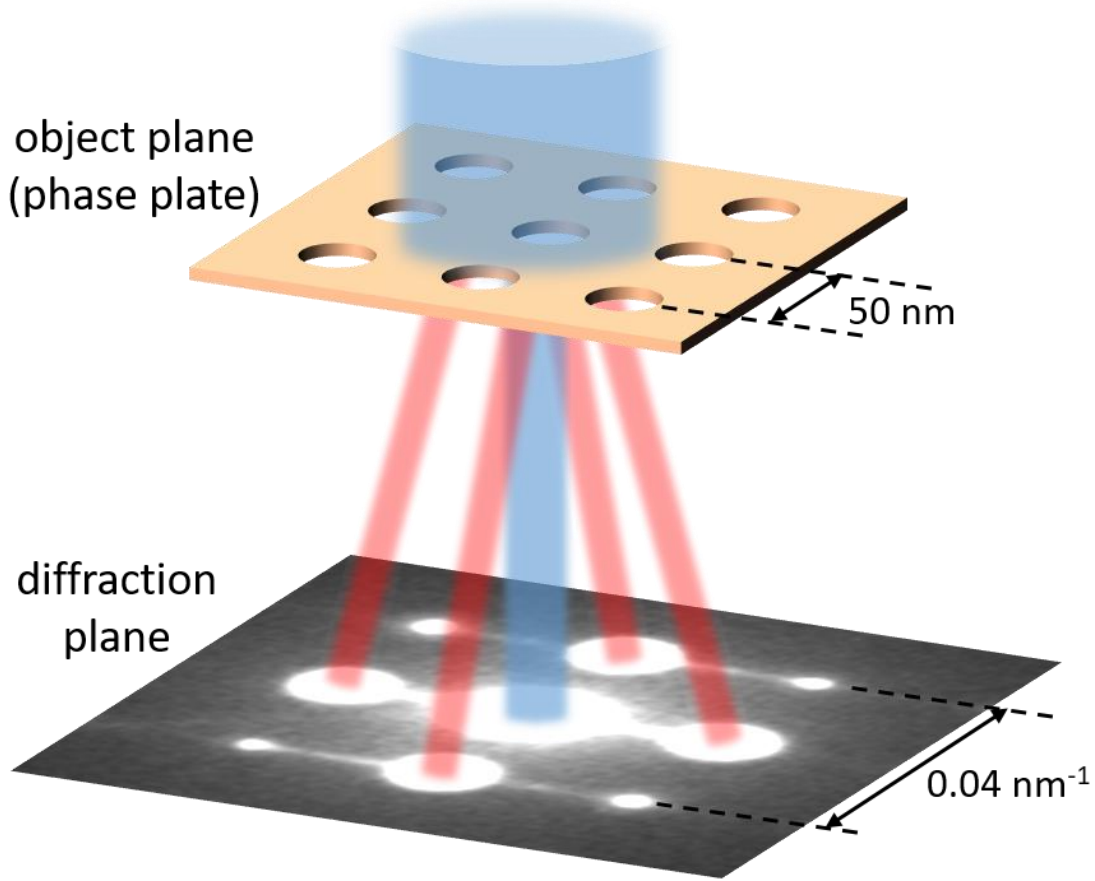

**Figure 1**. Schematic of electron diffraction from a nanostructured mesh phase plate. Direct beam (blue) and four first-order diffracted beams (red) are shown. The beam cross section image at the bottom is an experimentally measured electron diffraction pattern, recorded in a JEOL 2010F TEM using 200 keV electrons. For simplicity, the electron optics for focusing and imaging are not shown here.

The mesh phase plates were fabricated using electron-beam lithography and a reactive-ion-etching process (Figure 2a). We started from a silicon nitride membrane TEM grid (*SiMPore Inc.*) with 10 nm nominal membrane thickness. Electron-beam lithography was used to define the mesh pattern in the spin-coated poly(methyl methacrylate) (PMMA) thin film, a positive-tone electron-beam resist. The pattern was then transferred into the silicon nitride membrane via reactive-ion-etching to make through-holes in the membrane. After resist stripping, a metallic (gold or aluminum) film was evaporated onto the sample. The purpose of the metallization is twofold: (i) the metal film reduces charging effects in the electron-beam imaging and diffraction experiments; and (ii) this additional film gives us the ability to control the electron phase shift by varying the film composition and thickness. A more detailed description of the fabrication process can be found in the Methods section.

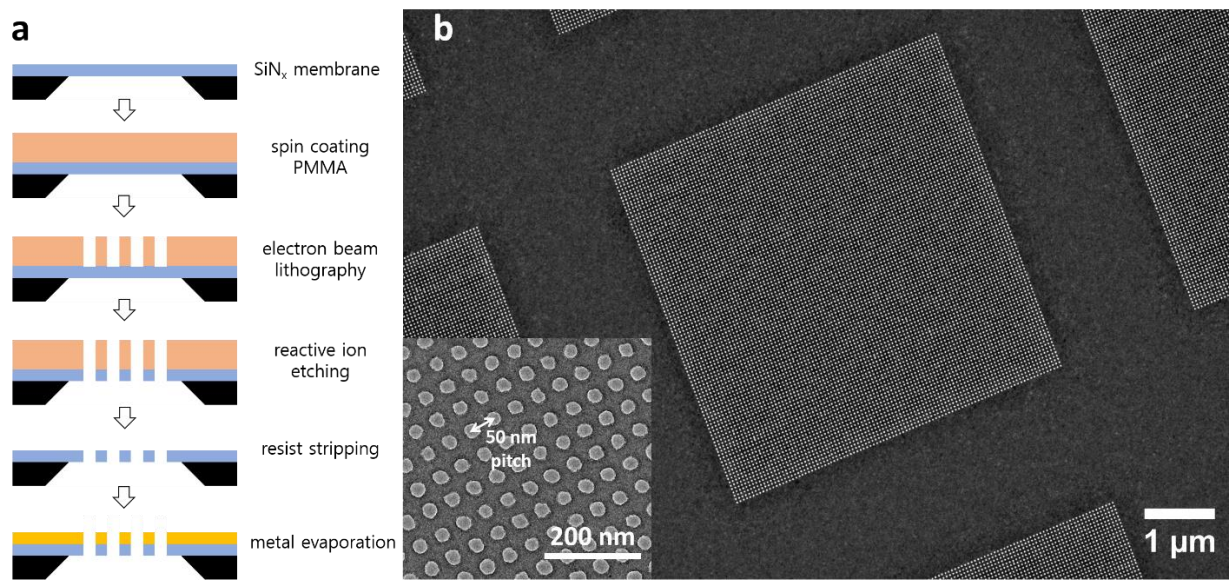

**Figure 2**. Large-area mesh nanostructures in a membrane. **a**, Fabrication process flow. **b**, A TEM image of the nanostructured membrane. Inset: a TEM image of the nanostructured membrane at a higher magnification. The image was taken in a Tecnai G2 TEM with 80 keV electrons.

Figure 2b shows a TEM image of a typical nanofabricated mesh phase plate. The nanoscale through-holes in the membrane form a square-lattice with 50 nm pitch. There are 2-μm-wide supporting bars in the horizontal and vertical directions to maintain the mechanical strength of the membrane (Figure S1 illustrates the function of the supporting bars). These supporting bars divide the nano-hole mesh into 5-μm-by-5-μm square regions. The patterned mesh covers a total area of 100-μm-by-100-μm on the membrane. We also fabricated nanostructured membranes with line grating patterns. However, for membrane gratings, the grating lines with a high aspect ratio tended to stick together, making it difficult to fabricate large area phase plates (Figure S2). The mesh phase plates were free from this issue (Figure S3).

The nanostructured electron phase plate was characterized with electron diffraction measurement using 200 keV electrons (JEOL 2010F TEM). Figure 3 shows the electron diffraction pattern as well as the high-dispersion electron diffraction pattern from the nanostructured mesh phase plate. A typical polycrystalline electron diffraction pattern was observed, with a central direct beam and several concentric rings. This diffraction pattern arose from the evaporated polycrystalline gold film. Given the pitch of the mesh nanostructure (50 nm) and the wavelength of 200 keV electrons (2.5 pm), the diffracted beams from the mesh phase plate should have a small angle (~50 μrad) and be located very close to the direct beam (~0.02 nm$^{-1}$). In order to observe the diffraction from the mesh phase plate, a high-dispersion electron diffraction pattern was taken (inset of Figure 3) in the low-magnification mode where the TEM objective lens was turned off. In the high-dispersion electron diffraction pattern, a square lattice of diffraction spots was observed in the reciprocal space, corresponding to the mesh nanostructure in the real space. The diffraction spots were labelled according to crystallography conventions. The distance between the diffraction spots (0.02 nm$^{-1}$) was also commensurate with the pitch (50 nm) of the mesh phase plate. The periodic mesh nanostructure can be considered as an "artificial crystal" that produces an electron diffraction pattern with a well-defined lattice structure.

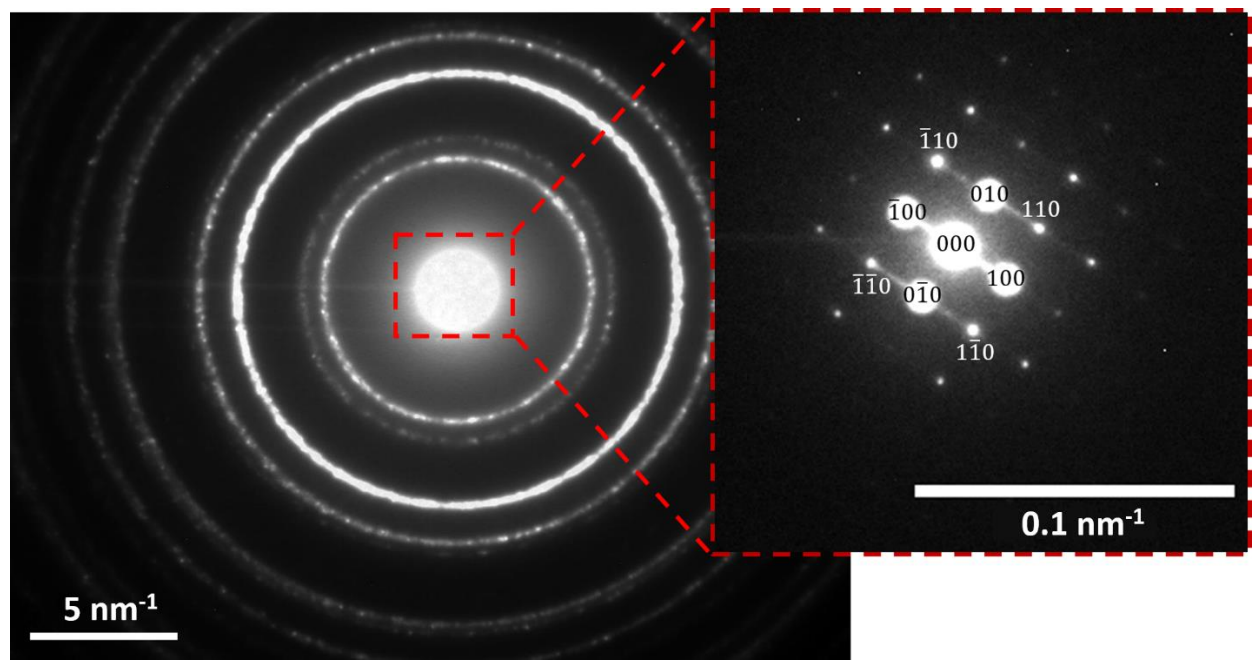

**Figure 3**. Electron diffraction patterns of the nanostructured membrane. The diffraction patterns were recorded in a JEOL 2010F TEM and the electron-beam energy was 200 keV. The diffraction pattern consists of a bright spot in the center representing the direct beam and a series of concentric rings coming from electron diffraction by the polycrystalline gold film. Inset: high-dispersion electron diffraction pattern of the nanostructured membrane. Note the focus and stigmation were re-adjusted to obtain the high-dispersion electron diffraction pattern. The square lattice in the reciprocal space corresponds to the mesh structure in the real space. The diffraction spots were labelled according to crystallography conventions.

Changing the phase profile of a phase plate can modify its diffraction pattern. We changed the phase profile of the mesh phase plate by adjusting the electron-beam energy, thus modulating the intensities of different diffraction orders. The electron diffraction experiment was conducted in a TEM (FEI Tecnai G2) with its beam energy adjustable from 20 keV to 120 keV. Figure 4 shows the TEM image of the mesh phase plate and its selected-area electron diffraction patterns at various electron-beam energies. A selected-area aperture selected a region on the mesh phase plate without supporting bars so that the diffraction came from only the mesh structure. In the diffraction patterns, the zeroth ({000}), first ({100}), and second ({110}) diffraction orders are most visible. The diffraction efficiency, namely the relative intensity of different diffraction orders, was modulated by the beam energy. The diffraction efficiency modulation was most obvious in the diffraction pattern from 60 keV electrons, where the direct beam was suppressed compared to first-order and second-order diffracted beams (Figure 4c). The modulation of diffraction intensities by changing the beam energy (and hence the phase shift) is analogous to oscillations of diffraction intensities with sample thickness (and hence the phase shift) in crystal diffraction. We interpret this diffraction intensity modulation as an effective Pendellosung effect[26] in an artificial nanostructure.

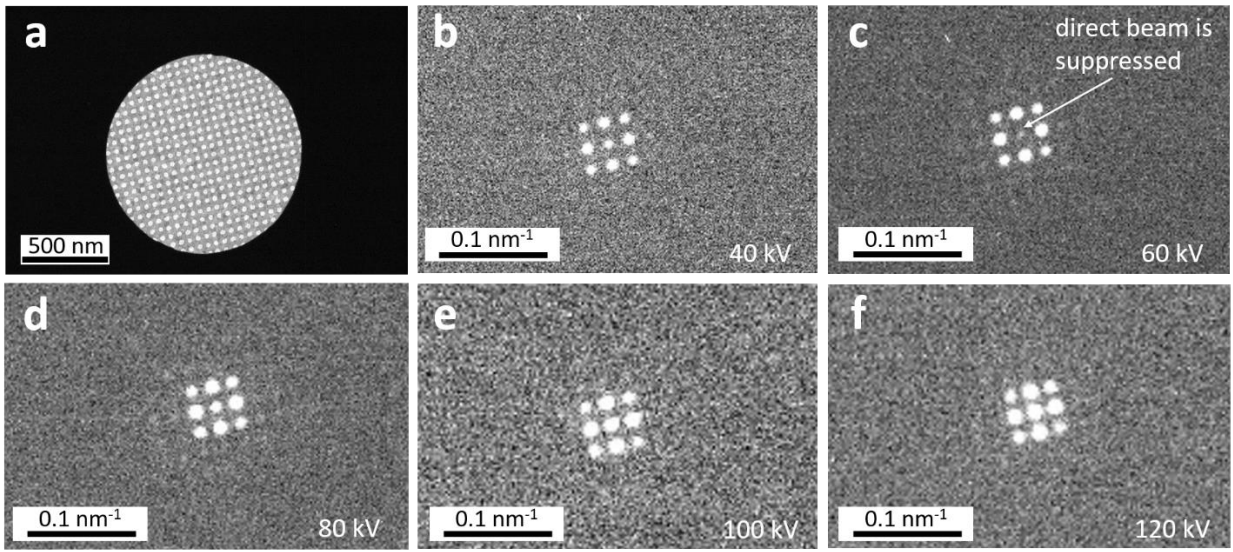

**Figure 4**. Selected-area electron diffraction patterns from the nanostructured membrane using electron beams with various energies. The diffraction patterns were recorded in a Tecnai G2 TEM. **a**, A TEM image of the nanostructured membrane and the selected-area aperture. **b-f**, Selected-area electron diffraction patterns with a 40 keV (**b**), 60 keV (**c**), 80 keV (**d**), 100 keV (**e**), and 120 keV (**f**) electron beam.

For a quantitative comparison of the mesh phase plate electron diffraction patterns at different electron energies, we measured the intensity ratio between the direct beam ($I_0$) and the first-order diffracted beam ($I_1$) (averaged from the four first-order diffracted beams: $(100)$, $(\bar{1}00)$, $(010)$, $(0\bar{1}0)$) (Figure 5a). It was confirmed that, by changing the electron energy, and hence the phase profile of the mesh phase plate, the intensities of diffracted beams could be modulated. The intensity ratio between the second-order diffracted beam ($I_2$) (averaged from the four second-order diffracted beams: $(110)$, $(\bar{1}10)$, $(1\bar{1}0)$, $(\bar{1}\bar{1}0)$) and the first-order diffracted beam ($I_1$) was also measured and shown in Figure 5a. Furthermore, the phase profile of the phase plate could be changed by using different material compositions. We fabricated mesh phase plates with an aluminum film instead of a gold film on top of the silicon nitride membrane. The selected-area electron diffraction patterns for various electron energies are shown in Figure S5. The intensity distribution among diffraction orders of an aluminum-coated mesh phase plate is different from that of a gold-coated mesh phase plate. For instance, the direct beam was never suppressed for all the electron energies tested. The measured electron diffraction beam intensity ratios ($I_0/I_1$ and $I_2/I_1$) for an aluminum-coated mesh phase plate are shown in Figure 5b.

The measured beam-intensity ratio was verified by theoretical modeling of electron diffraction from the mesh phase plate. The mesh phase plate was modeled as a uniform film with a periodic hole array in it. The mesh was considered as a mixed amplitude and phase plate. When a free electron passes through a thin film, the phase shift (relative to propagation in free space) can be described as[27]

$$\Delta\phi = C_E V_0 t = \frac{2\pi e(E + E_0)}{\lambda E(E + 2E_0)} V_0 t$$

where $C_E$ is a constant depending on electron energy, $V_0$ is the material mean inner potential (MIP), $t$ is the material thickness, e is the electron charge, $\lambda$ is the electron wavelength, $E$ is the electron kinetic energy, and $E_0$ is the electron rest energy. Thus, the phase plate is binary, with 0 phase shift in the holes and $\Delta\phi$ phase shift in the membrane structure. Note the $\Delta\phi$ phase shift is electron-energy dependent as well as material dependent. For the amplitude modulation, a constant transmission factor was applied to the electron wave passing through the membrane structure. The calculated beam-intensity ratios ($I_0/I_1$ and $I_2/I_1$) are shown in Figure 5. For the gold-coated phase plate (Figure 5a), the best fit between theoretical and experimental diffraction beam-intensity ratios was obtained by setting the MIP-thickness product to 260 V · nm and the amplitude transmission factor to 0.38 (more details can be found in the Methods section). The fitted MIP-thickness product appears smaller than the estimated value assuming 10 nm thickness for both Au and $Si_3N_4$, and 21 V MIP for Au[27] and 14~16.7 V MIP for $Si_3N_4$[8,28]. We attribute this discrepancy to the following potential causes: (i) beam-energy-dependent inelastic scattering and high-angle diffraction from the phase plate are not considered in the theoretical model; (ii) the actual material thicknesses are different from the nominal thicknesses; (iii) the material density and composition can vary depending on the fabrication conditions and material thickness (e.g. the Au film was only ~10 nm thick and could be different from the bulk Au, resulting in a lower MIP[29]); and (iv) the pattern of the fabricated phase plate can be different from the ideal circular hole arrays. Using the same parameters, the calculated beam-intensity ratio between the second-order diffracted beam ($I_2$) and the first-order diffracted beam ($I_1$) is also shown in Figure 5. In theory, the intensity ratio $I_2/I_1$ is a constant. This constant intensity ratio between diffracted beams is a result of the binary phase modulation imposed by the phase plate, and this constant only depends on the phase plate pattern and the diffraction orders, free from any fitting parameters. In experiments, the intensity ratio $I_2/I_1$ slightly increases with a decreasing electron energy. This increase could be caused by the fact that electrons with a lower energy are subject to a stronger scattering by the phase plate material, leading to a higher relative intensity of the second-order diffracted beams as they have a larger scattering angle. For the aluminum-coated mesh phase plate (Figure 5b), the best fit between theoretical and experimental diffraction beam-intensity ratios was obtained by setting the MIP-thickness product to 200 V·nm and the amplitude transmission factor to 0.68. Both a lower MIP-thickness product and a higher amplitude transmission factor are expected by changing the Au film to the Al film. The fitted MIP-thickness product also appears smaller than the estimated value assuming 10 nm thickness for both Al and $Si_3N_4$, and 13.9 V MIP for Al[30] and 14~16.7 V MIP for $Si_3N_4$[8,28]. The aluminum-coated mesh phase plates also present a constant intensity ratio $I_2/I_1$ that is well reproduced in the theory. Compared with gold-coated phase plates, the closer fitting of $I_2/I_1$ between theory and experiments could be a result of weaker energy-dependence of the electron scattering.

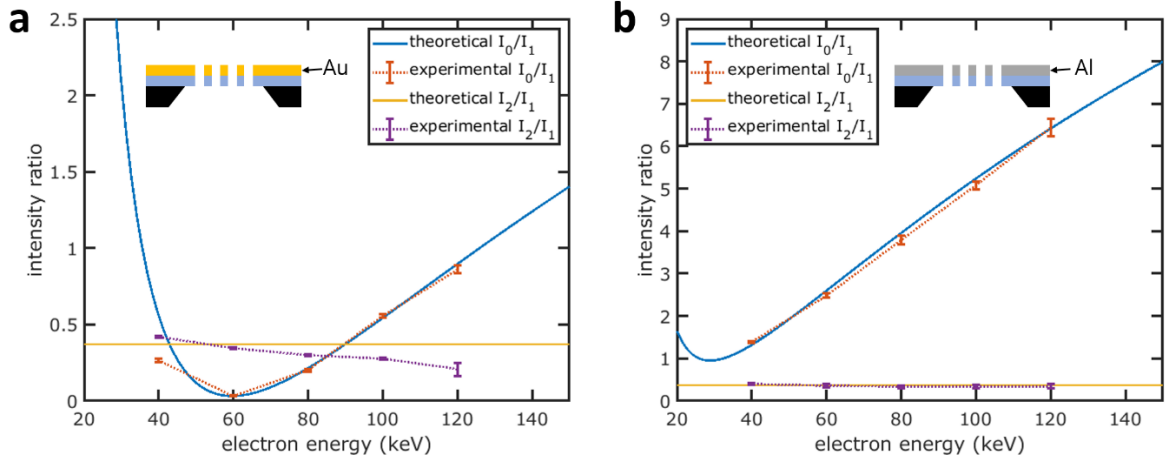

**Figure 5**. Experimental and theoretical beam-intensity ratios (the direct beam ($I_0$) to the first-order diffracted beam ($I_1$), and the second-order diffracted beam ($I_2$) to the first-order diffracted beam ($I_1$)) as a function of electron-beam energy. **a**, Au-coated phase plate. **b**, Al-coated phase plate. Each of the experimental data point was obtained by measuring the beam-intensity ratio from electron diffraction patterns taken at 8-20 (varies for different electron energies) different exposure times, with the error bars showing the standard deviations.

We further performed diffraction experiments using the large-area nanostructured phase plates to produce diffractive images in a scanning electron microscope (SEM)[31]. Figure 6a shows the experimental setup. The mesh phase plate is inserted in the beam path of the SEM and functions as a diffraction grating. The estimated beam spot size was 57 μm on the phase plate. The beam spot size was estimated by using the edges of the window in the silicon frame to perform a knife-edge measurement. While the SEM electron beam is scanning across the sample, multiple diffracted beams are also scanning across the sample and generating multiple superimposed images with small offsets. Figure 6b is a low magnification SEM image showing the sample to be imaged as well as the membrane TEM grid with the mesh phase plate (opaque region is the silicon frame and transparent region is the membrane). The sample contained Sn nanoparticles with various sizes, with the SEM image (without the diffraction grating) shown in Figure 6c. Figure 6d illustrates a diffractive image of the sample, showing multiple superimposed and displaced images of the nanoparticles. A fast Fourier transform (FFT) of the diffractive image is also shown, with its spatial frequency components resembling the electron diffraction pattern of the mesh phase plate.

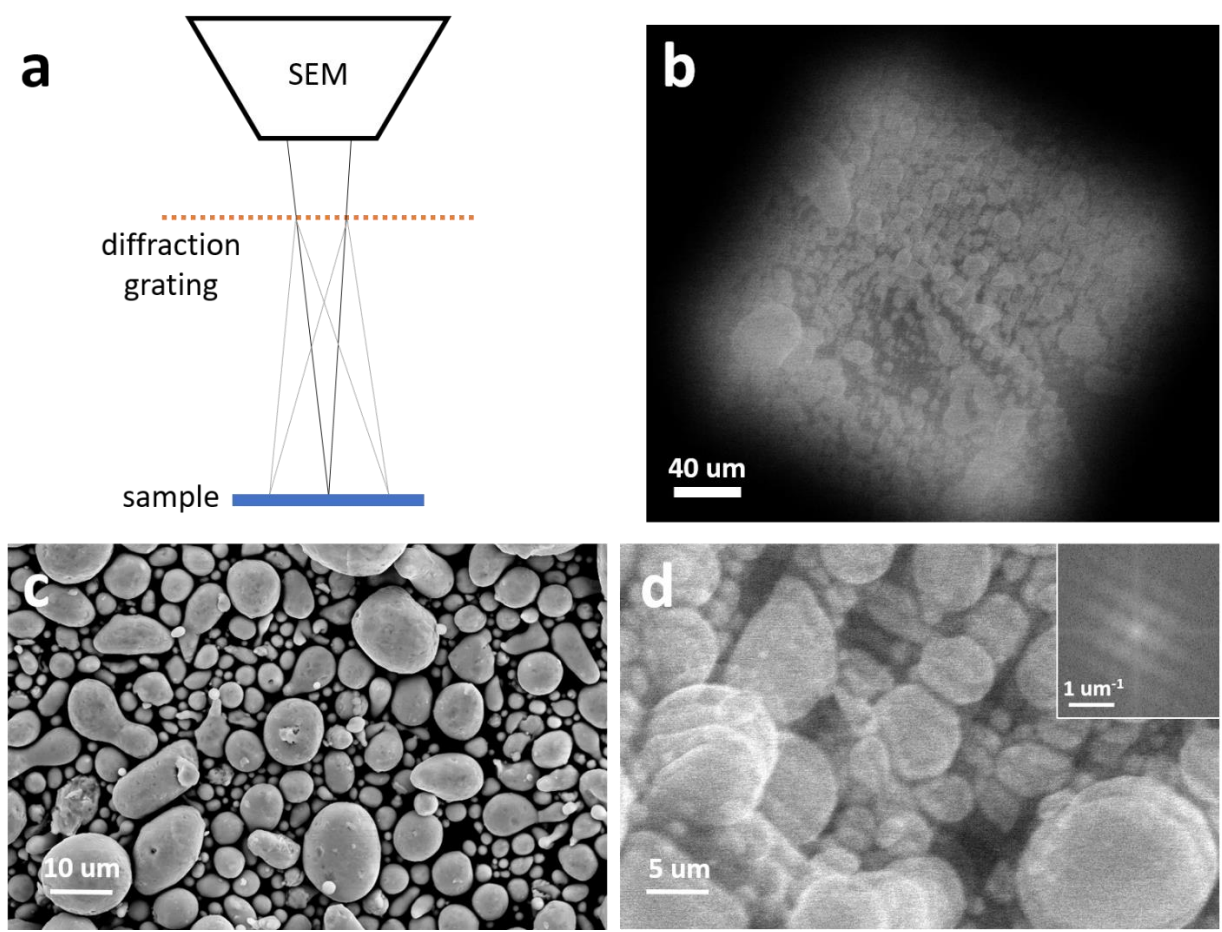

**Figure 6**. SEM diffractive imaging with the nanostructured membrane. **a**, Experimental setup of diffractive imaging. The nanostructured membrane diffraction grating is inserted into the beam

path between the SEM column and the sample. The SEM electron beam is diffracted by the membrane into multiple beams focused at the sample. As these beams scan across the sample, multiple superimposed and displaced images are generated. **b**, An SEM image of diffractive imaging of Sn nanoparticles, showing the nanostructured membrane (transparent square region) and its Si supporting frame (opaque region). The estimated beam spot size was 57 μm on the phase plate. The beam spot size was estimated by using the edges of the window in the silicon frame to perform a knife-edge measurement. **c**, A regular SEM image (without the diffraction grating) of Sn nanoparticles used as the sample for diffractive imaging. **d**, A diffractive SEM image of the Sn nanoparticles. Inset: fast Fourier transform (FFT) of the image.

In conclusion, we demonstrated large-area electron phase plates fabricated on an electron-transparent membrane with electron-beam lithography and reactive-ion-etching. We characterized the phase plate with electron diffraction in TEMs as well as diffractive imaging in an SEM. We observed that the phase plates created a null in the center of the bright-field for certain electron energy. The nullification of the direct beam was explained by the coherent interference of diffracted beams. This special beam-shaping capability could potentially be used in a variety of dark-field electron-microscopy techniques to improve the spatial resolution, increase the image contrast, and boost the signal-to-noise ratio. The large-area electron phase plates could also enable electron-wavefront-engineering in an SEM, as shown by the diffractive imaging. Furthermore, our work showed that the functionality of electron phase plates, such as the diffraction efficiency, could be tuned by changing the electron energy. The static nature of electron phase plates usually renders their application inconvenient, as a new phase plate has to be fabricated and swapped into the electron beam path whenever a new functionality is required, or if the desired functionality is compromised by fabrication imperfections. For instance, the diffraction efficiency of electron phase plates was shown to be sensitive to the surface profiles of the membrane nanostructures[9,19]. Recently, a programmable phase plate with micro-fabricated electrostatic-elements has been shown to achieve tunable pixelated phase modulation[26]. Here, we demonstrated that tuning the electron phase plates was also achieved by adjusting the electron energy, which could provide an alternative route towards the active tuning of electron phase plates, and the creation of electron phase plates with adjustable functions.

**Acknowledgements**

This work was supported by Gordon and Betty Moore Foundation. We gratefully acknowledge helpful discussions with Prof. Pieter Kruit and Prof. James LeBeau. We thank Akshay Agarwal and Navid Abedzadeh for assistance with editing the manuscript. We would also like to thank Dr. Vitor Manfrinato, Jim Daley and Mark Mondol for assistance with sample fabrication.

**Methods**

**Sample fabrication.** Silicon nitride membrane TEM grids (*SiMPore Inc.*) were used for sample fabrication. Each membrane had a nominal thickness of $10 \pm 0.5$ nm (from vendor specifications) and was supported by a silicon frame with 250-μm-by-250-μm windows. A 25 s oxygen plasma ashing was used to clean the grid and promote resist adhesion. The grid was then stuck to a sacrificial silicon chip with carbon tapes for spin-coating. A PMMA thin film, a positive-tone electron-beam resist, was spin-coated with approximately 70 nm thickness. The nanostructure pattern was defined by electron-beam lithography via an Elionix F-125 system using 125 keV electrons. The exposure dose was 4800 ~ 6400 µC/cm$^2$ (300 ~ 400 e$^-$/nm$^2$). After exposure, cold development was performed at 0°C in 3:1 IPA:MIBK. CF$_4$ reactive-ion-etching transferred the pattern into the membrane by etching through-holes. After etching, PMMA resist was stripped via a 90 s oxygen plasma ashing. Metallization was performed by e-beam evaporation of gold or aluminum with 10 nm nominal thickness (the nominal thickness was achieved by measuring the deposition rate with a quartz crystal microbalance and setting the deposition time corresponding to a 10-nm-thick film).

**Electron diffraction measurement.** High-dispersion electron diffraction was performed in a JEOL 2010F TEM with 200 keV electrons (see Figure 3). In this diffraction mode, the objective lens was switched off. The electron beam illuminated a full window in the silicon nitride TEM grid containing the patterned mesh nanostructure. The electron diffraction pattern was recorded with a camera length of 80 m. Selected-area electron diffraction was performed in a Tecnai G2 TEM with a tunable electron energy (20-120 keV) (see Figure 4). In order to resolve the diffraction pattern from the mesh phase plate, the incident electron beam was spread out to ensure a small convergence angle and hence small spots in the diffraction pattern. The region under illumination was selected via an aperture (see Figure 4a). The electron diffraction patterns were recorded with a camera length of 4.2 m. The beam intensity was measured from the diffraction pattern by first subtracting the background and then integrating the intensity around the corresponding diffraction spot. To minimize the artifacts of the TEM image recording system, we took the same diffraction pattern with several different exposure times. For one phase plate at one electron energy, the beam-intensity ratio should be constant. We only used diffraction patterns with an intermediate exposure time and they produced similar beam-intensity ratios. Diffraction patterns with a low exposure time was susceptible to noise, while diffraction patterns with a high exposure time suffered from camera saturation. At the intermediate exposure time, between 8 and 20 electron diffraction patterns (varies for different electron energies) were used to produce the average value and the standard deviation of each experimental data point showing the beam intensity ratios in Figure 5.

**Electron diffraction theory.** The mesh phase plate was modeled as a uniform film with a periodic circular hole array in it. Thus, the phase plate is binary, with 0 phase shift in the holes and $\Delta\phi$ phase shift in the membrane structure. The phase shift $\Delta\phi$ is calculated via

$$\Delta\phi = C_E V_0 t = \frac{2\pi e(E + E_0)}{\lambda E(E + 2E_0)} V_0 t$$

The mesh was considered as a mixed amplitude and phase plate. For the amplitude modulation, a constant transmission factor was applied to the electron wave passing through the membrane structure. The electron diffraction pattern was calculated as the Fourier transform of the mesh phase plate. The analytical Fourier coefficients are

$$c_{m,n} = \alpha e^{i\Delta\phi} \frac{\sin(m\pi)}{m\pi} \frac{\sin(n\pi)}{n\pi} + (1 - \alpha e^{i\Delta\phi}) \frac{\pi d^2}{2p^2} \frac{J_1(\pi d\sqrt{m^2 + n^2}/p)}{\pi d\sqrt{m^2 + n^2}/p}$$

where integer pair $(m, n)$ denotes the diffraction order, $\alpha$ is the amplitude transmission factor, $d$ is the hole diameter, $p$ is the array pitch, and $J_1$ is the Bessel function of the first kind with order 1. The diffraction intensity is hence $I_{m,n} = |c_{m,n}|^2$. The Fourier transform was also calculated numerically, producing the same results as the analytical analysis. The hole diameter $d = 28$ nm was measured from a typical TEM image of the phase plate by averaging over ~200 nano-holes. To fit the experimental diffraction data from a gold-coated phase plate, the MIP-thickness product $V_0 t$ and the amplitude transmission factor $\alpha$ were varied, and the best fit was obtained for $V_0 t = 260$ V·nm and $\alpha = 0.38$. The fitted MIP-thickness product appears smaller than the estimated value assuming 10 nm thickness for both Au and Si$_3$N$_4$, and 21 V MIP for Au[27] and 14~16.7 V MIP for Si$_3$N$_4$[8,28]. This discrepancy is attributed to nonidealities in phase plate fabrication and the absence of inelastic scattering and high angle diffraction in the theoretical model. To fit the experimental diffraction data from an aluminum-coated phase plate, the best fit was obtained for $V_0 t = 200 \text{ V} \cdot \text{nm}$ and $\alpha = 0.68$ (due to the almost linear behavior of the experimental data, similar goodness of fit was also obtained by using a slightly smaller MIP-thickness product and a slightly larger amplitude transmission factor, or using a slightly larger MIP-thickness product and a slightly smaller amplitude transmission factor). Both a lower MIP-thickness product and a higher amplitude transmission factor are expected by changing the Au film to the Al film. The fitted MIP-thickness product also appears smaller than the estimated value assuming 10 nm thickness for both Al and Si$_3$N$_4$, and 13.9 V MIP for Al[30] and 14~16.7 V MIP for Si$_3$N$_4$[8,28].

**SEM diffractive imaging.** Diffractive imaging was performed in a modified Zeiss LEO 1525 SEM. The electron energy was 20 keV (see Figure 6). A PELCO STEM imaging holder (*Ted Pella Inc.*) was used to hold both the nanostructured membrane phase plate and the Sn nanoparticle sample with ~18 mm separation. The electron beam was focused at the Sn nanoparticles and diffractive imaging was acquired from the secondary electron signal at the Everhart-Thornley detector of the SEM. The beam spot size on the phase plate was estimated by using the edges of the window in the silicon frame as knife edges (Figure 6b), and measuring the image grayscale along a line-scan across the knife edge.

24

**Supporting Information:**

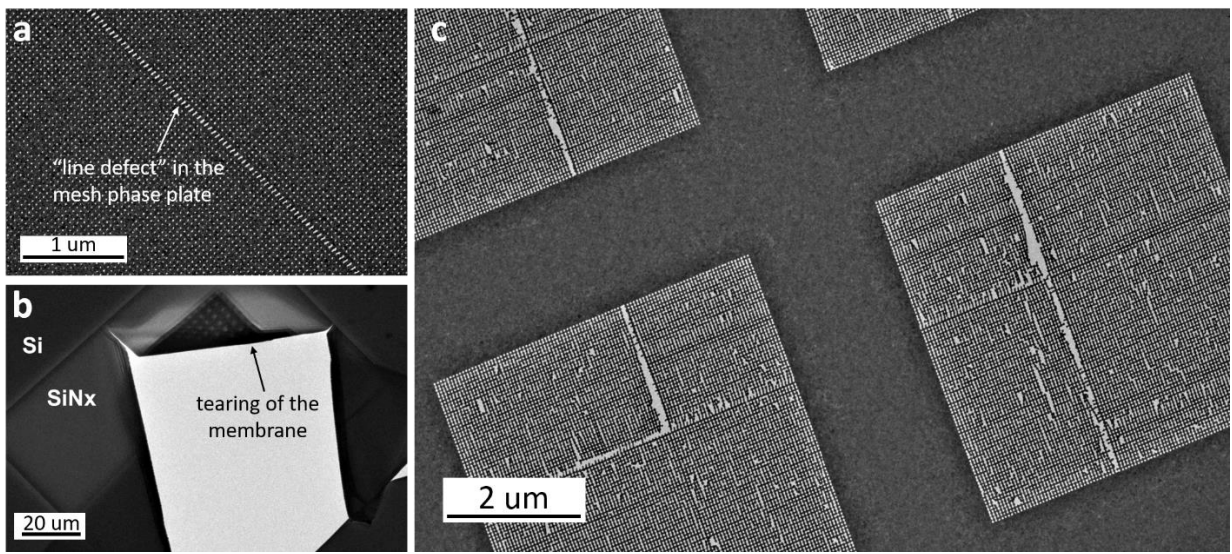

**Figure S1**. TEM images showing supporting bars maintain the mechanical strength of the nanostructured membrane. **a,** Defects in the nanostructured membrane can occur along a line. **b,** Without supporting bars, these defects eventually lead to a tearing of the nanostructured membrane. **c.** With supporting bars, the defects are confined within the patterned area, and the membrane is not broken even with the presence of defects.

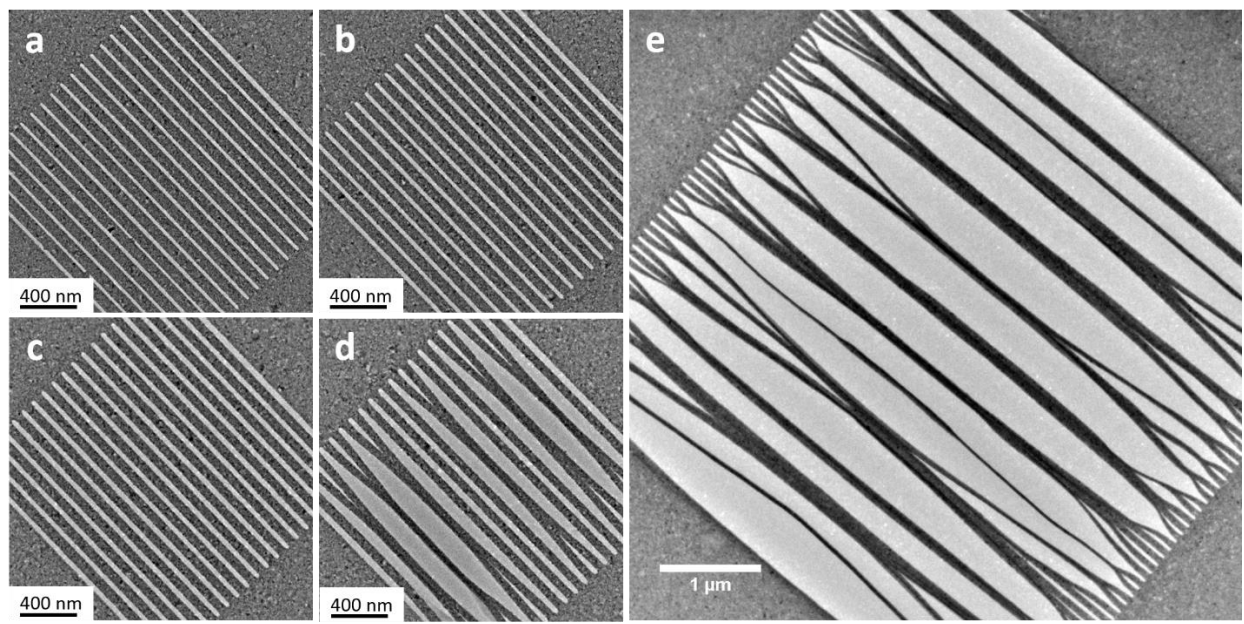

**Figure S2.** TEM images of line grating nanostructures in a membrane. **a-d,** Gratings with 2-μm-long lines and various widths and hence aspect ratios. For the grating with narrow and high aspect ratio lines (**d**), the adjacent grating lines tend to stick together. **e,** A larger area grating with 6-μm-long lines. The grating lines have a high aspect ratio and adjacent lines tend to stick together.

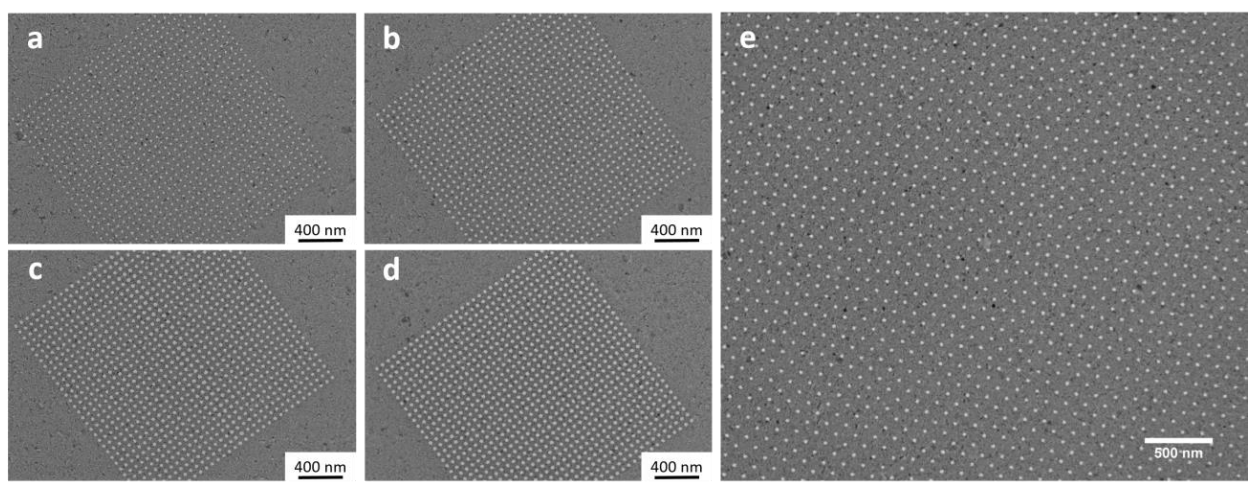

**Figure S3.** TEM images of mesh nanostructures in a membrane. **a-d,** 2-µm-by-2-µm meshes with various sized nano-holes. **e,** A larger area 6-µm-by-6-µm mesh. In contrast to line gratings, the meshes are free from defects.

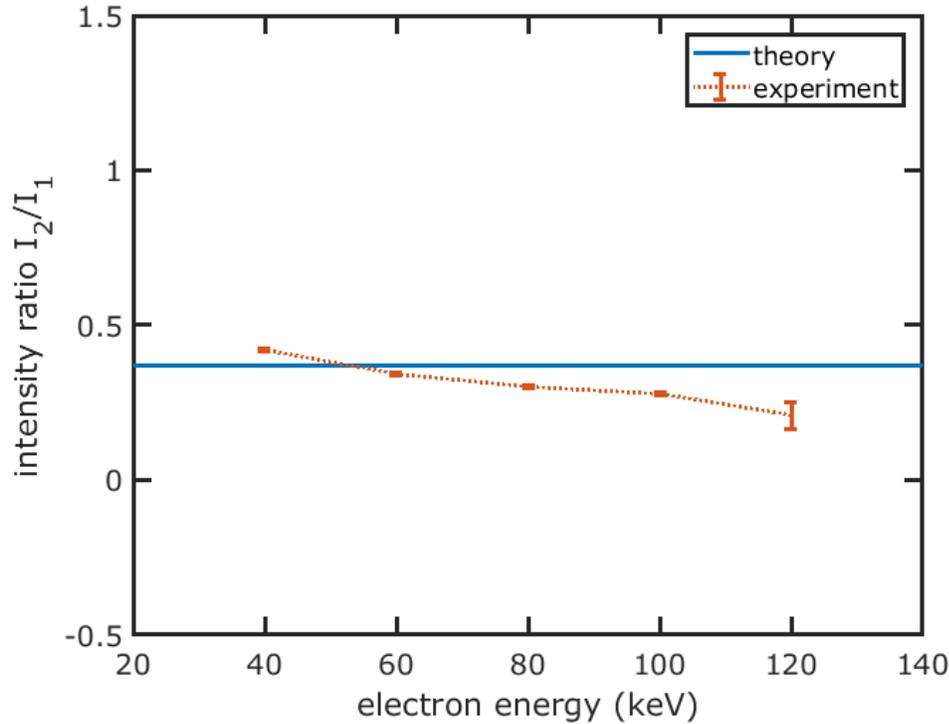

**Figure S4.** Experimental and theoretical beam-intensity ratio between the second-order diffracted beam ($I_2$) and the first-order diffracted beam ($I_1$) as a function of electron-beam energy. Each of the experimental data point was obtained by measuring the beam-intensity ratio from electron diffraction patterns taken at 8-15 (varies for different electron energies) different exposure times, with the error bars showing the standard deviations. In theory, the intensity ratio $I_2/I_1$ is a constant, since the phase plate impose a binary phase modulation. In experiments, the intensity ratio $I_2/I_1$ slightly increases with a decreasing electron energy. This increase could be caused by the fact that electrons with a lower energy are subject to a stronger scattering by the phase plate material,

leading to a higher relative intensity of the second-order diffracted beams as they have a larger scattering angle.

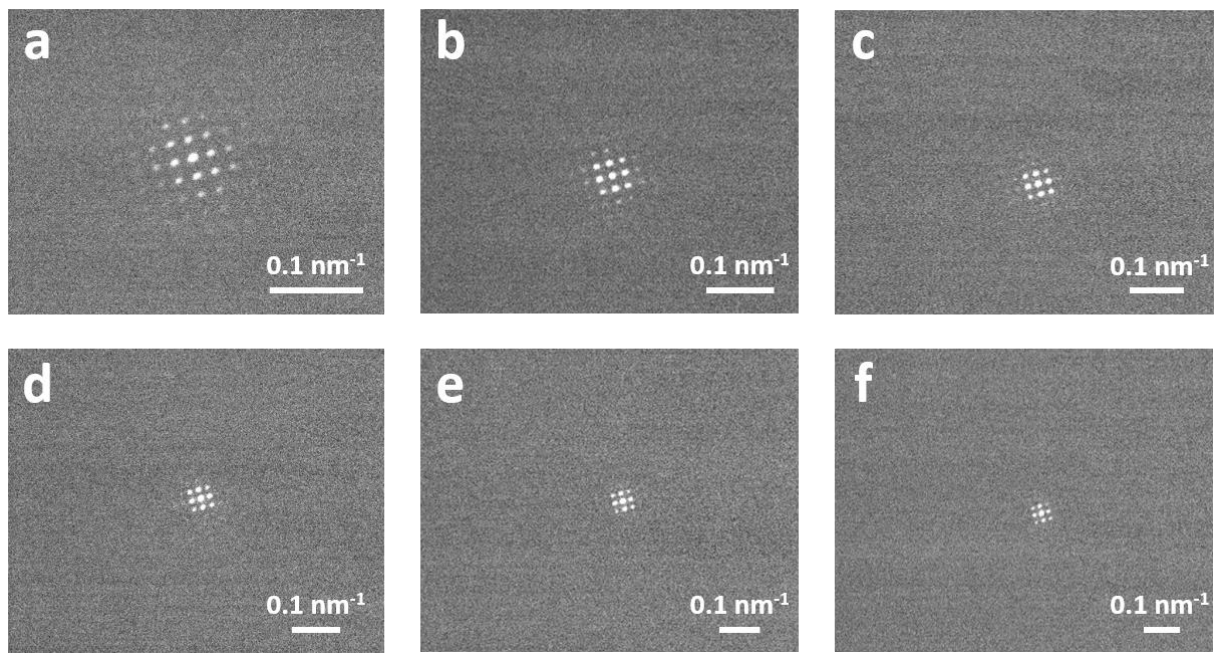

**Figure S5**. Selected-area electron diffraction patterns from an Al-coated nanostructured membrane. The electron-beam energy is 20 keV (**a**), 40 keV (**b**), 60 keV (**c**), 80 keV (**d**), 100 keV (**e**), and 120 keV (**f**).